\documentclass[twocolumn]{aastex62}
\usepackage{enumitem}
\graphicspath{{./}{figures/}}
\usepackage[T1]{fontenc}
\usepackage{epsfig}
\usepackage{epstopdf}
\usepackage{natbib}
\usepackage{amssymb}
\usepackage{amsbsy}
\usepackage{natbib}
\usepackage{subfigure}
\usepackage[mathcal]{euscript}
\usepackage{float}
\usepackage{amsmath}
\usepackage{tabularx}
\usepackage{xspace}
\usepackage{enumitem}
\usepackage{xcolor}
\usepackage{comment}
\usepackage{hyperref}
\usepackage{ulem}

\usepackage{ulem,xspace}

\newcommand{\msun}{$M_{\odot}$}

\newcommand{\logg}{\mbox{$\log g$}}

\newcommand{\ctwelvecthirteen}{$^{12}$C/$^{13}$C\xspace}

\newcommand{\grad}{\ensuremath{\nabla}}
\newcommand{\gradrad}{\ensuremath{\nabla_{\rm{rad}}}}
\newcommand{\gradad}{\ensuremath{\nabla_{\rm{ad}}}}
\newcommand{\gradmu}{\ensuremath{\nabla_{\mu}}}

\newcommand{\Dth}{\ensuremath{D_\mathrm{th}}}
\newcommand{\Numu}{\ensuremath{\mathrm{Nu}_\mu}}

\bibpunct{(}{)}{;}{a}{}{,}

\begin{document}

\title{Characterizing Observed Extra Mixing Trends in Red Giants using the \textit{Reduced Density Ratio} from Thermohaline Models}


\author[0000-0003-4323-2082]{Adrian E. Fraser}
\altaffiliation{These authors contributed equally to this work}
\affiliation{Department of Astrophysical and Planetary Sciences \& LASP, University of Colorado, Boulder, CO 80309, USA}
\affiliation{Department of Applied Mathematics, Baskin School of Engineering, University of California, Santa Cruz, CA 95064, USA}
\affiliation{Kavli Institute for Theoretical Physics, University of California, Santa Barbara, CA 93106, USA}

\author[0000-0002-8717-127X]{Meridith Joyce}
\altaffiliation{These authors contributed equally to this work}
\affiliation{Konkoly Observatory, Research Centre for Astronomy and Earth Sciences, H-1121 Budapest Konkoly Th. M. \'ut 15-17., Hungary}
\affiliation{CSFK, MTA Centre of Excellence, Budapest, Konkoly Thege Mikl\'os \'ut 15-17., H-1121, Hungary}
\affiliation{Lasker Fellow, Space Telescope Science Institute,
3700 San Martin Drive,
Baltimore, MD 21218, USA}
\affiliation{Kavli Institute for Theoretical Physics, University of California, Santa Barbara, CA 93106, USA}

\author[0000-0002-3433-4733]{Evan H. Anders}
\altaffiliation{These authors contributed equally to this work}
\affiliation{CIERA, Northwestern University, Evanston IL 60201, USA}
\affiliation{Kavli Institute for Theoretical Physics, University of California, Santa Barbara, CA 93106, USA}

\author[0000-0002-4818-7885]{Jamie Tayar}
\altaffiliation{These authors contributed equally to this work}
\affiliation{NASA Hubble Fellow}
\affiliation{Department of Astronomy, University of Florida, Bryant Space Science Center, Stadium Road, Gainesville, FL 32611, USA }
\affiliation{Institute for Astronomy, University of Hawai‘i at Mānoa, 2680 Woodlawn Drive, Honolulu, HI 96822, USA}
\affiliation{Kavli Institute for Theoretical Physics, University of California, Santa Barbara, CA 93106, USA}

\author[0000-0001-5048-9973]{Matteo Cantiello}
\affiliation{Center for Computational Astrophysics, Flatiron Institute, New York, NY 10010, USA}
\affiliation{Department of Astrophysical Sciences, Princeton University, Princeton, NJ 08544, USA}
\affiliation{Kavli Institute for Theoretical Physics, University of California, Santa Barbara, CA 93106, USA}

\correspondingauthor{Adrian Fraser}
\email{Adrian.Fraser@colorado.edu}
\correspondingauthor{Meridith Joyce}
\email{meridith.joyce@csfk.org}
\correspondingauthor{Evan Anders}
\email{evan.anders@northwestern.edu}
\correspondingauthor{Jamie Tayar}
\email{jtayar@ufl.edu}

\begin{abstract}
Observations show an almost ubiquitous presence of extra mixing in low-mass upper giant branch stars. The most commonly invoked explanation for this is thermohaline mixing. One-dimensional stellar evolution models include various prescriptions for thermohaline mixing, but the use of observational data directly to discriminate \textit{between} thermohaline prescriptions has thus far been limited. Here, we propose a new framework to facilitate direct comparison: Using carbon-to-nitrogen measurements from the SDSS-IV APOGEE survey as a probe of mixing and a fluid parameter known as the \textit{reduced density ratio} from one-dimensional stellar evolution programs, we compare the observed amount of extra mixing on the upper giant branch to predicted trends from three-dimensional fluid dynamics simulations. Using this method, we are able to empirically constrain how mixing efficiency should vary with the reduced density ratio. We find the observed amount of extra mixing is strongly correlated with the reduced density ratio and that trends between reduced density ratio and fundamental stellar parameters are robust across choices for modeling prescription. We show that stars with available mixing data tend to have relatively low density ratios, which should inform the regimes selected for future simulation efforts. Finally, we show that there is increased mixing at low reduced density ratios, which is consistent with current hydrodynamical models of thermohaline mixing. The introduction of this framework sets a new standard for theoretical modeling efforts, as validation for not only the amount of extra mixing, but trends between the degree of extra mixing and fundamental stellar parameters is now possible.
\end{abstract}


\section{Introduction} 
\label{sec:intro}
\setcounter{footnote}{0}
\subsection{Astrophysical Context}
Observations of globular clusters and low-metallicity field stars show significant changes in the abundance ratios of elements known to be sensitive to mixing, including \ctwelvecthirteen, lithium, and [C/N], as a star evolves up the red giant branch \citep{Carbon1982, Pilachowski1986, Kraft1994, Shetrone2019}. These changes occur around the red giant branch bump (RGBB) and are largest in the most metal-poor stars \citep[e.g.][]{Gratton2000}. Large samples of stars that can be used to trace this mixing are now available from a variety of spectroscopic surveys, including GALAH \citep{buder2019}, APOGEE \citep{DR17}, and GAIA-ESO \citep{Magrini2021b}. However, observed surface abundance trends are in tension with standard theoretical stellar evolution models, which predict that the surface chemistry should not evolve in this regime.

As low-mass stars ascend the red giant branch, they undergo a series of mixing and homogenizing events as their interior burning and energy transport zones interact. Near the base of the red giant branch, the surface convection zone reaches its deepest level of penetration into the stellar interior, leaving behind a chemical discontinuity from which it recedes in subsequent evolution. This inflection in the convection zone's movement is known as the ``first dredge-up.'' The red giant branch bump occurs when the outward-propagating hydrogen burning shell encounters this chemical discontinuity, triggering a structural realignment in which the star's core contracts and the luminosity drops, causing a disruption to the otherwise monotonic increase in luminosity along the red giant branch \citep{Christensen-Dalsgaard:2015}. In one dimensional (1D) stellar models, the sensitivity of the RGBB to physical assumptions makes it a powerful diagnostic of interior mixing processes \citep[e.g.][]{FusiPecci1990,Salaris2002, Joyce2015, Khan2018}. 
However, in standard models of red giant stars, there is no mixing between the hydrogen-burning shell and the overlying convective envelope after the first dredge-up, and no change in surface abundances is predicted in this regime. This is in direct conflict with abundance trends found in observations.

The most widely studied candidate mechanism for rectifying this discrepancy is thermohaline mixing, driven by an inversion of the mean molecular weight $mu$ caused by $^3$He burning. 
The potential for $^3$He burning to cause a $\mu$ inversion and drive thermohaline mixing (though not specifically in low-mass RGB stars) was first pointed out by \citet{Ulrich1972}. Later, this $\mu$ inversion was identified as a significant source of mixing in low-mass RGB stars \citep{Dearborn2006, Eggleton2006, Eggleton2007}, but was connected to the Rayleigh-Taylor instability, not thermohaline mixing. The specific connection between extra mixing in low-mass RGB stars and \textit{thermohaline mixing} was later made by \citet{charbonnel_thermohaline_2007}. As the hydrogen-burning shell moves into the region chemically homogenized by the  first dredge-up, the $^3$He($^3$He, 2p)$^4$He reaction creates an inversion of the mean molecular weight $\mu$. While this $\mu$ inversion is insufficient to generate a convective region (c.f. \citealt{CantielloLanger2010}), these conditions give rise to the 
\textit{thermohaline instability}, a phenomenon perhaps best known in the context of salt water in Earth's oceans \citep{Stern1960,baines_gill_1969}. 

\subsection{Fluid Dynamics Context} \label{intro:subsec:fluids}
Thermohaline mixing occurs in Ledoux-stable regions that have stably stratified temperature gradients but unstable mean molecular weight stratification 
(see, e.g., \citealt{garaud_DDC_review_2018} for a full review or \citealt{SalarisCassisi2017} for discussion focused on a 1D stellar modeling context).
Thermohaline mixing is a double-diffusive phenomenon present in fluids that have different diffusivities for heat and chemical composition that, in turn, make opposing contributions to the radial density gradient \citep{Turner:1974}.
This process may facilitate the radial mixing of elements between the hydrogen-burning shell and the stellar convective envelope, thus producing measurable changes in the surface mixing diagnostics after the first dredge-up. 

Fluid dynamicists have studied thermohaline mixing in great detail, often employing 3D simulations in ``local'' domains, i.e., periodic boundary conditions with constant linear background gradients in temperature $T$ and mean molecular weight $\mu$ under the Boussinesq approximation \citep{spiegel_boussinesq_1960}. Within this standard framework, the efficiency of thermohaline mixing, $\Dth$ (the degree to which thermohaline mixing enhances chemical mixing via turbulent motions, as explained in Sec.~\ref{sec:parameterizations} below), strictly depends on three dimensionless numbers, which we introduce here but describe in further detail in Secs.~\ref{sec:formalism} and \ref{sec:parameterizations}. 
The Prandtl number $\mathrm{Pr}$ and diffusivity ratio $\tau$ characterize molecular diffusivities and are fluid properties, meaning they are independent of background gradients. 
$\mathrm{Pr}$ is the ratio of the kinematic viscosity to thermal diffusivity, and $\tau$ is the ratio of the compositional and thermal diffusivities. In stars, $\mathrm{Pr}\approx\tau \ll 1$, so double-diffusive instabilities like thermohaline mixing are readily driven \citep{garaud_DDC_review_2018}. 

The third dimensionless quantity describing thermohaline mixing is the density ratio $R_0$, which is the ratio of the temperature gradient's stabilizing contribution to the density divided by the $\mu$ gradient's destabilizing contribution to the density. The density ratio is a measure of how conducive to driving thermohaline mixing the the conditions in a radial location within a star are. 
If the density ratio is large, then the destabilizing $\mu$ gradient is weak and the system may be stable (if $R_0$ exceeds a threshold that depends on $\tau$) or only weakly unstable to the thermohaline instability. A small density ratio, on the other hand, corresponds to a significant inversion of the $\mu$ profile and thus the system is strongly unstable to thermohaline mixing and possibly even convection. 

In this work, we study the \emph{reduced density ratio}, $r$, which combines $R_0$ and $\tau$ into a single quantity that directly determines the instability of a system to the thermohaline instability \citep{traxler_etal_2011}. The reduced density ratio directly defines the fluid dynamical stability of a system such that
\begin{equation}
r
    \begin{cases}
    \leq 0 & \mbox{System is convectively unstable} \\
    \in (0,1) & \mbox{System is thermohaline unstable} \\
    \geq 1 & \mbox{System is stable}
    \end{cases}.
\end{equation} 
We stress that $r$ is \emph{not} a measure of the \emph{efficiency} or mixing rate of thermohaline mixing. Rather, it is a measure of the structural stability of a system, or its tendency to mix. While mixing efficiency, $\Dth$, does depend on $r$, the exact dependence is an open question. Several simplified mixing prescriptions exist for predicting how efficiency depends on different fluid parameters \citep[see review by][]{garaud_DDC_review_2018}. 
As we will show in Sec.~\ref{sec:parameterizations}, often these mixing prescriptions diverge significantly in their predictions for how mixing efficiency ($\Dth$) varies with $r$.

\subsection{Open Questions}
Given that the physical conditions required to trigger the thermohaline instability are in place at around the same time that extra mixing has been observed in red giant stars \citep[e.g.][]{Lagarde2015}, most authors have assumed that all of the observed extra mixing can be attributed to the thermohaline instability \citep[e.g.][]{Kirby2016, Charbonnel2020, Magrini2021a}. However, this connection has also been questioned for a number of reasons. 

First, reproducing the observed amounts of mixing in this regime with 1D models requires 
assuming that thermohaline mixing is much more efficient than most fluid simulations would suggest is reasonable \citep{Denissenkov2010thermohaline, denissenkov_merryfield_2011, traxler_etal_2011, brown_etal_2013}. Questions have likewise been raised about whether the evolutionary timing of the observed extra mixing is truly consistent with thermohaline models \citep[see e.g.][]{Angelou2015, Henkel2017, TayarJoyce22}.

Additionally, authors have put forth many different prescriptions for including thermohaline mixing in 1D models. 
Many previous works (for example, \citealt{CharbonnelLagarde2010}, \citealt{Lagarde2017}) have used observations to calibrate or constrain 1D stellar model parameters that control the overall mixing efficiency within a particular mixing prescription (for example, $\alpha_{\text{th}}$ in the Kippenhahn prescription, described in Sec.~\ref{sec:parameterizations} below).  
These efforts generally employ a single prescription \citep[e.g.~that of][]{kippenhahn_etal_1980} and calibrate that prescription to observations. 
However, calibrating free parameters (e.g. $\alpha_{\text{th}}$, see Sec.~\ref{sec:parameterizations}) within any given mixing prescription, while useful in a 1D context, does not allow us to discriminate between the different proposed mixing prescriptions, which predict different (even, in some cases, opposite) trends in mixing efficiency $\Dth$ versus fluid parameters like $r$.
We are unaware of any instance in which observations of mixing have been used to (in)validate proposed theoretical mixing prescriptions. 

This issue is all the more pressing in light of recent work demonstrating that models of thermohaline instability that include the presence of a relatively low-amplitude magnetic field can result in much more efficient mixing (larger $\Dth$) and \emph{significantly different dependence of that efficiency on $r$} \citep{harrington}. 
Thus, while the issue that 1D models need to assume extremely efficient mixing in order to agree with observations could potentially be addressed by the inclusion of magnetic effects, it is not clear whether the profoundly different trend in mixing efficiency versus $r$ is consistent with observations.

With more observational data available in stars across different masses and metallicities, a natural question to ask is: can these data be used to identify which of the different relationships between mixing efficiency ($\Dth$) and $r$ are more consistent with observations? This question is challenging because observations show how the amount of mixing mixing varies with mass and metallicity, while different mixing prescriptions relate mixing efficiency to $r$. Fluid properties (e.g.~$\mathrm{Pr}$, $\tau$) are readily extracted from 1D stellar evolution models \citep[e.g.][]{Jermyn_Anders_atlas}, thus allowing values to be inferred for stars across different masses and metallicities. 
In contrast, we are unaware of any prior work which has extracted $r$ from 1D stellar evolution models, which is required in order to understand how observed mixing trends relate to thermohaline models developed from 3D dynamical simulations.  

\subsection{Purpose of the present study}
Given these observational and theoretical questions, the development of a framework through which we can determine whether particular thermohaline prescriptions are more or less consistent with observations 
is timely and imperative. 
In this paper, we put forth such a framework: one that 
shifts the focus away from
the calibration of individual mixing efficiency parameters within particular 1D thermohaline mixing prescriptions, and instead facilitates inter-comparison among the mixing models themselves. 
\begin{quote}
We demonstrate a robust and model-agnostic means of relating the non-dimensional fluid parameters relevant to thermohaline mixing (particularly $r$) to the observed mixing around the RGB bump and show that this correlation is indeed qualitatively consistent with 1D prescriptions of thermohaline mixing informed by 3D simulations. 
\end{quote}
Further, while previous work \citep[e.g.][]{charbonnel_thermohaline_2007} has used the measurements of the \textit{overall amount} of extra mixing to tune model parameters that control the \textit{overall efficiency} of thermohaline mixing prescriptions, our framework allows us to use trends in extra mixing as a function of fundamental stellar parameters to probe trends in mixing efficiency ($\Dth$) as a function of fluid parameters.
Using this framework, we demonstrate that the magnetized thermohaline mixing prescription put forth by \citet{harrington}, which addressed significant outstanding problems by enhancing mixing efficiency over hydrodynamic levels, is not consistent with observations (when fixing their magnetic parameter $H_B$).

This paper is organized as follows: we begin by summarizing the formalism and stellar structure quantities relevant to thermohaline mixing (Sec.~\ref{sec:formalism}). This is followed by a description of various 1D mixing prescriptions commonly adopted in stellar evolution calculations (Sec.~\ref{sec:parameterizations}). We then introduce a suite of 1D MESA simulations and calculate the relevant fluid parameters in the thermohaline region for a range of mass and metallicity assumptions (Secs.~\ref{sec:mesa_experiment} and \ref{sec:mesa_results}). Finally, we compare an observational proxy of extra mixing, the decrease in [C/N] near the RGBB, to theoretical trends predicted by existing 1D thermohaline mixing prescriptions (Sec.~\ref{sec:obs} and Sec.~\ref{sec:punchline}). Our results and their implications are discussed in Sections \ref{sec:punchline} and \ref{sec:conclusions}. 

\section{Thermohaline Formalism } 
\label{sec:formalism}
The following section presents a derivation of the fluid parameter known as the ``reduced density ratio,'' which is a fundamental parameter in fluid models of thermohaline mixing, and measures the stability of a given fluid configuration to the instability that drives thermohaline mixing. 
Section \ref{sec:parameterizations} goes into a similar level of detail regarding different prescriptions for thermohaline mixing in 1D stellar models 
to the end of summarizing the prescriptions currently implemented in MESA, and demonstrating key differences between these prescriptions. As was noted in Sec.~\ref{intro:subsec:fluids}, while the amount of chemical mixing predicted by these prescriptions does depend on the density ratio (or, equivalently, the reduced density ratio), we stress that the density ratio characterizes stability and thus the tendency to mix, not the actual amount of mixing or mixing efficiency (which is characterized by $\Dth$, introduced in Sec.~\ref{sec:parameterizations}).

The instability driving thermohaline mixing requires a Ledoux-stable inversion of the mean molecular weight $\mu$ stratification in the presence of a stable temperature gradient. 
The stability of the temperature gradient is given by the Schwarzschild criterion:

\begin{equation} \label{eq:Schwarzschild}
    \gradrad - \gradad < 0,
\end{equation}
where the temperature gradient $\grad \equiv d \ln P / d \ln T$ (pressure $P$ and temperature $T$) has an adiabatic value $\grad = \gradad$ and saturates to $\grad = \gradrad$ in hydrostatically stable regions where the flux is carried radiatively. 
The Ledoux criterion for convective stability is \citep{Ledoux1947}

\begin{equation} \label{eq:Ledoux}
    \gradrad - \gradad - \frac{\phi}{\delta}\gradmu < 0,
\end{equation}
which must be satisfied despite the inversion of the mean molecular weight.
The Ledoux criterion accounts for the composition gradient $\gradmu = d\ln\mu/d\ln P$, where $\delta = -(\partial \ln \rho / \partial \ln T)_{P,\mu}$ and $\phi = (\partial \ln \rho / \partial \ln\mu)_{P,T}$ (where $\rho$ is density).

To be unstable, the destabilizing $\mu$ gradient (the free energy source that drives thermohaline mixing) must be sufficiently strong relative to the stabilizing influence of the temperature gradient. This is quantified by the density ratio,
\begin{equation} \label{eq:R0}
    R_0 \equiv \frac{\grad - \gradad}{\frac{\phi}{\delta} \gradmu},
\end{equation}
where $R_0 < 1$ implies the $\mu$ gradient is sufficiently unstable to drive convection, and $R_0 > 1$ implies the fluid is stably-stratified (i.e.~no convection)\footnote{Note that $R_0>1$ is equivalent to the Ledoux criterion Eq.~\eqref{eq:Ledoux} only if $\grad = \gradrad$. Thermohaline mixing primarily mixes chemicals, but does produce some minimal thermal mixing \citep[see, e.g., Fig.~4 of][]{brown_etal_2013}; thus, $\grad \neq \gradrad$. This thermal mixing is often ignored in mixing prescriptions in 1D stellar evolution programs, however.}. 
As reviewed by \citet{garaud_DDC_review_2018}, fluids with $R_0 > 1$ can be prone to double-diffusive instabilities whenever the thermal diffusivity, $\kappa_T$, is greater than the compositional diffusivity, $\kappa_\mu$. Specifically, the instability driving thermohaline mixing acts whenever

\begin{equation} \label{eq:R0_condition}
1 < R_0 < 1/\tau,
\end{equation}
\citep{baines_gill_1969} where
\begin{equation} \label{eq:tau}
    \tau \equiv \kappa_\mu/\kappa_T.
\end{equation}
Note that in stellar radiation zones, typically $\tau \lesssim 10^{-6}$. This means that very slight inversions of $\mu$ (large $R_0$) can drive thermohaline mixing, even when the temperature gradient is strongly stable according to the Schwarzschild criterion. 

Throughout this paper, we express the density ratio $R_0$ in terms of the \textit{reduced density ratio}

\begin{equation} \label{eq:r}
    r \equiv \frac{R_0 - 1}{\tau^{-1} - 1}
\end{equation}
per, e.g., \citet{traxler_etal_2011,brown_etal_2013}.
In terms of $r$, the condition for thermohaline instability, Eq.~\eqref{eq:R0_condition}, is

\begin{equation} \label{eq:r_condition}
    0 < r < 1,
\end{equation}
where $r \leq 0$ is the threshold for convection and $r \geq 1$ corresponds to scenarios where the $\mu$ inversion is too weak to drive the thermohaline instability.

\section{Parameterized Thermohaline Models } 
\label{sec:parameterizations}
Equation \eqref{eq:r_condition} 
can be readily evaluated at any radial location in a model star generated with a 1D stellar structure and evolution program. However, predicting the efficiency of thermohaline mixing is much more challenging. A diffusive approximation is commonly taken for the turbulent mixing of chemicals such that the total mixing of chemical species is given by the sum of the molecular diffusivity and a turbulent mixing coefficient, $\Dth$. This diffusion coefficient, $\Dth$, relates the rate of chemical mixing to the chemical composition gradient, and is broadly what is meant when referring to ``mixing efficiency" throughout this paper. This quantity can be converted to the compositional Nusselt number discussed in the fluid dynamics literature, $\Numu$, using the formula

\begin{equation} \label{eq:Dth_from_Nu}
    \Dth = (\Numu - 1)\kappa_\mu.
\end{equation}

We call any model that predicts $\Dth$ as a function of stellar structure variables (e.g.~gradients and molecular diffusivities of chemicals and heat) a parameterized mixing model or mixing prescription. 
Efforts to develop thermohaline mixing prescriptions for use in models of stellar interiors date back many decades, see \citet{garaud_DDC_review_2018} for a full review. 
Such mixing prescriptions have been implemented in a variety of 1D stellar evolution programs \citep[see][and references therein]{lattanzio_etal_2015}, enabling studies of the effects of thermohaline mixing in stars across the Hertzprung-Russell diagram. 
Here, we briefly summarize the most commonly used and more recently developed prescriptions.

The \textit{de facto} thermohaline mixing model used in MESA (first described in \citealt{CantielloLanger2010} and implemented for public use in \citealt{mesa2}) is commonly referred to as the ``Kippenhahn model'' and was originally derived by \citet{Ulrich1972} and \citet{kippenhahn_etal_1980}.
Using arguments based on dimensional grounds and assumptions about the shapes and motions of discrete fluid parcels, they derived a mixing efficiency of the form

\begin{equation} \label{eq:Dth-kipp}
    \Dth = C_t \kappa_T R_0^{-1},
\end{equation}
\citep[see Eq.~(5) of][]{charbonnel_thermohaline_2007}
where $C_t$ is a free parameter, with plausible values ranging from $C_t = 658$ \citep{Ulrich1972} to $C_t = 12$ \citep{kippenhahn_etal_1980}. 
We note that Eq.~\eqref{eq:Dth-kipp} predicts finite mixing for $r \geq 1$ ($R_0 \geq 1/\tau$), even though thermohaline mixing is formally stabilized for these parameters.

\begin{figure}
    \centering
    \includegraphics[width=\columnwidth]{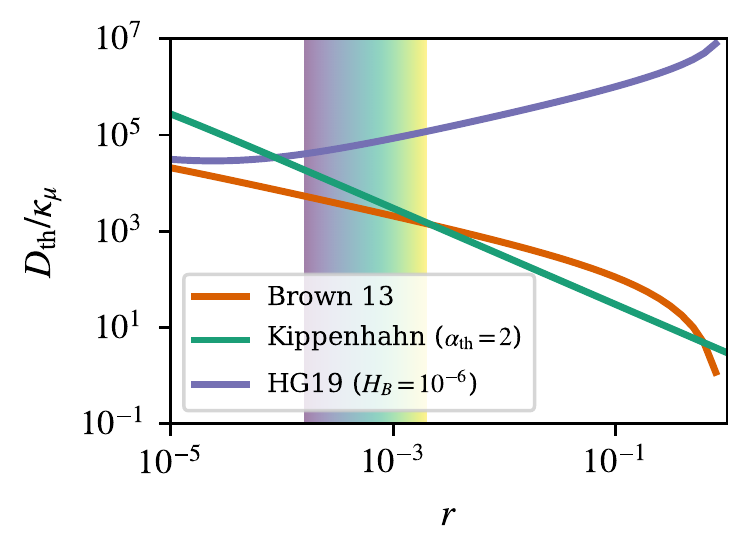}
    \caption{ 
    Prescriptions of the compositional diffusivity due to thermohaline mixing $D_{\rm th}$ normalized by the molecular diffusivity $\kappa_{\mu}$ are plotted against the reduced density ratio $r$. For each prescription, we use $\mathrm{Pr} = \tau = 10^{-6}$, consistent with the conditions in these regions of RGB stars.
    We plot two hydrodynamic models, the \citet{brown_etal_2013} model (orange) and the \citet{kippenhahn_etal_1980} model with $\alpha_{\rm th} = 2$ (green). In both cases, the mixing efficiency decreases with $r$.
    The \citet{harrington} model (HG19) is also shown; it includes magnetic fields, which cause mixing efficiency to increase with $r$ for these parameters.
    The plotted curve for the HG19 model depends on $H_B$, which depends on the stellar structure and magnetic field strength; the plotted value is characteristic of the structure in the thermohaline zone of a 1.1 $M_\odot$ star at [Fe/H] = -0.2 with a magnetic field whose strength is $\mathcal{O}(100 \,\mathrm{G})$.
    The purple-to-yellow color gradient plotted in the background denotes the range of $r$ values that we measure in our grid of 1D stellar evolution models, which are displayed in Fig.~\ref{fig:mesa_r_spread}.
    }
    \label{fig:parameterization_compare}
\end{figure}

Nevertheless, Eq.~\eqref{eq:Dth-kipp} is implemented in MESA as
\begin{equation} \label{eq:Dth-kipp-MESA}
    \Dth = \frac{3}{2} \alpha_{\rm{th}} \frac{K}{\rho C_P}R_0^{-1}
\end{equation}
\citep[see Eq.~(14) of][]{mesa2}. 
Here, $\alpha_{\rm{th}}$ is a dimensionless efficiency parameter related to $C_t$ by $C_t = 3\alpha_{\rm{th}}/2$, $K$ is the radiative conductivity, $\rho$ is the density, and $C_P$ is the specific heat at constant pressure, with $\kappa_T = K/(\rho C_P)$. 
The green curve in Fig.~\ref{fig:parameterization_compare} shows $\Dth/\kappa_\mu$ vs.~$r$ calculated according to Eq.~\eqref{eq:Dth-kipp-MESA} for $\tau = 10^{-6}$, which is a representative value for the thermohaline-unstable region of RGB stars, and the same $\alpha_{\rm{th}} = 2$ assumed in \citet{CantielloLanger2010}.

In addition to tension regarding the choice of model parameters (e.g.~$\alpha_\mathrm{th}$) controlling overall mixing efficiency within a given 1D prescription \citep[see e.g.][for further discussion]{Ulrich1972, kippenhahn_etal_1980, charbonnel_thermohaline_2007, CantielloLanger2010,traxler_etal_2011}, there have also been questions about the appropriate trends in mixing efficiency as a function of fluid parameters (i.e.~how $\Dth$ should depend on quantities like $r$ and $\mathrm{Pr}$) and therefore the stellar structure variables on which they depend \citep{garaud_DDC_review_2018}.
Thus, recent work has sought to refine these mixing prescriptions by performing numerical experiments with multi-dimensional simulations to more accurately parameterize mixing efficiency \citep{Denissenkov2010thermohaline,traxler_etal_2011}. 
\citet{traxler_etal_2011} and \citet{brown_etal_2013} performed 3D hydrodynamic simulations across a broad range of parameters. 
Not only did they find orders of magnitude less mixing than what is predicted by the Kippenhahn model with the model parameter required in \citet{charbonnel_thermohaline_2007} to find agreement with observations ($C_t = 1000$), they also developed new mixing prescriptions that fit their simulations much more closely. 
In the case of \citet{traxler_etal_2011}, the authors derived a mixing law by fitting an
analytic function 
of the form

\begin{equation} \label{eqn:trax_model}
   \Dth = \kappa_{\mu}\sqrt{\frac{\mathrm{Pr}}{\tau}}\left(a e^{-br}[1 - r]^c\right),
\end{equation}
to their simulation results,
where 

\begin{equation} \label{eq:Prandtl}
    \mathrm{Pr} = \frac{\nu}{\kappa_T}
\end{equation}
is the Prandtl number, with $\nu$ the kinematic viscosity,
and $a$, $b$, and $c$ are constants which they fit to data. 

While \citet{traxler_etal_2011} clearly showed their simulations are inconsistent with the mixing efficiency $\Dth$ implied by the Kippenhahn model with $\alpha_{\rm{th}}, C_t \sim 10^2-10^3$, it is important to note that their simulations generally explored $\mathrm{Pr}, \tau \sim 10^{-1}$, whereas these fluid parameters are generally of the order $10^{-6}$ in the radiative interiors of RGB stars. 
Thus, a fair question is whether mixing efficiency might increase to these larger values as $\mathrm{Pr}$ and $\tau$ approach $10^{-6}$. 
However, \citet{traxler_etal_2011} varied these parameters by an order of magnitude in their simulations, and investigated trends in $\Dth$. They found that mixing should not increase in this fashion, as indicated by the dependence of $\Dth$ on $\mathrm{Pr}$ and $\tau$ in Eq.~\eqref{eqn:trax_model}, which makes an argument that these models can be made to fit the observational data difficult to justify. 

\citet{brown_etal_2013} note that the model in Eq.~\eqref{eqn:trax_model} performs well at high $R_0$, but underestimates mixing at low $R_0$, particularly in the stellar regime of low Pr and $\tau$.
They develop a semi-analytical model for thermohaline mixing,

\begin{equation}
    \Dth = \kappa_{\mu}C^2\frac{\lambda^2}{\tau l^2(\lambda + \tau l^2)},
    \label{eqn:brown_model}
\end{equation}
where $\lambda$ is the growth rate of the fastest-growing linearly unstable mode, $l$ is its horizontal wavenumber, and $C \approx 7$ was fit to data from 3D hydrodynamic simulations.
Both $\lambda$ and $l$ are functions of $\mathrm{Pr}$, $\tau$, and $R_0$, and are obtained by finding the roots of a cubic and quadratic polynomial (their Eqs.~19 and 20).
The orange curve in Fig.~\ref{fig:parameterization_compare} shows $\Dth/\kappa_\mu$ vs.~$r$ calculated according to Eq.~\eqref{eqn:brown_model} for $\mathrm{Pr} = \tau = 10^{-6}$, representative values for the thermohaline-unstable regions of RGB stars. 
Note that $\Dth/\kappa_\mu \to 0$ as $r \to 1$ as expected, since the thermohaline instability becomes stable for $r \geq 1$.
We see that Eq.~\eqref{eq:Dth-kipp-MESA} with $\alpha_{\rm{th}} = 2$ agrees with this prescription for some values of $r$, suggesting that significantly larger values of $\alpha_{\rm{th}}$ are not consistent with 3D hydrodynamic simulations. 
While the general dependence of $\Dth/\kappa_\mu$ on $r$ is significantly different between these two models, they do both feature monotonically decreasing values of $\Dth/\kappa_\mu$ versus $r$. 
This prescription is implemented in MESA and has since been used in \citet{bauer_bildsten_2019} and other works. 

\citet{harrington} extended the work of \citet{brown_etal_2013} by performing 3D magnetohydrodynamic (MHD) simulations of thermohaline mixing with initially uniform, vertical magnetic fields of various strengths to approximate the effects of magnetic fields from external processes including, for instance, a global dipole field or a large-scale magnetic field left behind by a dynamo acting in the receding convective envelope. 
They found that magnetism strictly increases mixing efficiency, sometimes dramatically.
They developed a mixing prescription that accounts for this effect by building on the model of \citet{brown_etal_2013}.
The strength of the magnetic field is introduced into their model through their parameter $H_B$, which is proportional to the square of the magnetic field strength and depends on other stellar structure variables \citep[see Eq.~19 of][]{harrington}.
Their mixing prescription is of the form

\begin{equation} \label{eq:harrington_model}
    \Dth = \kappa_{\mu}K_B\frac{w_f^2}{\tau (\lambda + \tau l^2)},
\end{equation}
where $w_f$ is obtained by solving a quartic polynomial that includes the magnetic field strength through $H_B$, and $K_B \simeq 1.24$ is directly related to the constant $C$ in Eq.~\eqref{eqn:brown_model}.

This mixing prescription agreed remarkably well with their 3D simulations, which were limited to $r = 0.05$ but ranged in magnetic field strength over several orders of magnitude.
The prescription, which has not yet been implemented in MESA at the time of this writing, has two asymptotic limits, one where $\hat{w}_f^2 \propto B_0^2$ when the magnetic field strength $B_0$ is large, and one which reduces to the model of \citet{brown_etal_2013} when $B_0$ is small.

The purple curve in Fig.~\ref{fig:parameterization_compare} shows $\Dth/\kappa_\mu$ vs.~$r$ calculated according to Eq.~\eqref{eq:harrington_model} for the same parameter choices as the orange curve, and with $H_B = 10^{-6}$, appropriate for the thermohaline zone of a 1.1 $M_\odot$ star at [Fe/H] = -0.2 and a magnetic field whose strength is $\mathcal{O}(100 \,\mathrm{G})$. 
Note that this magnetic field strength dramatically increases mixing efficiency relative to the hydrodynamic values, particularly at larger values of $r$, whereas the model predicts the same mixing as the Brown model for $r \lesssim 10^{-5}$. 
For larger values of $r$, the dependence of $\Dth/\kappa_\mu$ on $r$ is profoundly different than either of the hydrodynamic models, with $\Dth/\kappa_\mu$ increasing with $r$, even as the thermohaline instability approaches marginal stability as $r \to 1$.

The dramatic enhancement in mixing efficiency predicted by this model for magnetic field strengths of even $\mathcal{O}(100\,\mathrm{G})$ presents a promising resolution to the tension discussed above, namely that 1D stellar evolution models can only reproduce observations by assuming that mixing is far more efficient than what is seen in 3D hydrodynamic simulations. 
However, while their prescription may predict mixing efficiencies that are comparable in overall magnitude to that of the Kippenhahn model with $C_t \sim 10^3$, the two prescriptions yield qualitatively different trends in $\Dth$ vs $r$.
Given the variance of the predictions of these models, we focus in this paper on showing how observations can be used to suggest the \textit{trends} in mixing that models should hope to explain rather than on trying to calibrate model parameters controlling the overall mixing \textit{efficiency} for a particular model, as has been done before, in order to provide a framework in which we can distinguish between mixing prescriptions.


\section{Stellar Evolution Models}
\label{sec:mesa_experiment}
We use MESA stable release version 21.12.21 to conduct 1D numerical simulations of stars incorporating the effects of thermohaline mixing for metallicities ranging from [Fe/H] $= -1.4$ to $0.4$ ($Z = 0.00068$ to $0.038$) and masses from 0.9 to 1.7 $M_{\odot}$ at resolutions of 0.2 dex and 0.2 $M_{\odot}$, respectively. We adopt the solar abundance scale of \citet{GrevesseSauval1998} and the corresponding opacities of \citet{IglesiasRogers1996}, with low-temperature opacities of \citet{Ferguson2005}. Our models assume a helium abundance and helium-enrichment ratio of $Y=0.2485$ and $\frac{dY}{dZ} = 1.3426$, respectively, as in \citet{tayar_etal_2022}. 
We use an Eddington T-$\tau$ relation for the atmospheric surface boundary conditions.
We adopt the mixing length theory (MLT) prescription of \citet{Cox1980} with a fixed value of $\alpha_{\text{MLT}}= 1.6$ times the pressure scale height ($H_p$). We use the Ledoux criterion for convective stability and neglect the effects of convective overshoot \citep{Ledoux1947}. 
We use the \verb|pp_extras.net| nuclear reaction network, which contains 12 isotopes. 
More details are available in Appendix \ref{app:mesa}, and the exact configuration of our physical and numerical parameter choices is available on the associated GitHub repository\footnote{\texttt{github.com/afraser3/Empirical-Magnetic-Thermohaline}}. MESA history files for each of the 200 simulations discussed in Sec.~\ref{sec:mesa_results} are available on Zenodo \citep{thermohaline_zenodo}.

Simulations are evolved at $1.25\times$ the default mesh (structural) resolution and $2\times$ the default time resolution on the pre-main sequence and main sequence. Once the models ascend the red giant branch and reach a surface gravity $\log g \le 3$, resolutions are increased to $2\times$ the default spatial resolution and $10\times$ the default temporal resolution, respectively. Optimal resolution values were determined according to the convergence tests detailed in Appendix \ref{app:resolution_test}.

We study four grids of stellar evolution simulations with different thermohaline mixing prescriptions. One grid employs the \citet{brown_etal_2013} prescription\footnote{Although the Brown prescription contains no free parameters in their original conception, a multiplicative factor on $\Dth$ has been introduced in the MESA implementation. Brown's model is reproduced by assigning the quantity \texttt{thermohaline\_coeff} = 1, as was done here}. 
while the other three employ the \citet{kippenhahn_etal_1980} prescription with coefficients $\alpha_{\rm th} \in [0.1, 2, 700]$.
The Kippenhahn $\alpha_{\rm th} = 0.1$ model has inefficient mixing and represents a regime where thermohaline mixing is present but weak.
We study $\alpha_{\rm th} = 2$ because (1) it is consistent with the default implementation in MESA and discussion in the instrument paper \citep{mesa2}; (2) it is used in previous work \citep{CantielloLanger2010, TayarJoyce22}; and (3) it is consistent with findings from 2D and 3D hydrodynamical simulations in stellar regimes 
\citep{Denissenkov2010, traxler_etal_2011, brown_etal_2013}. We also study a more traditional $\alpha_{\rm th} = 700$, which is of the order of literature values calibrated 
using stellar observations \citep{lattanzio_etal_2015, charbonnel_thermohaline_2007}.
We note that the timestep size of our $\alpha_{\rm th} = 700$ cases violate the CFL timestep constraint detailed by \citet{lattanzio_etal_2015}; MESA implicitly timesteps diffusive terms and so this violation does not affect the stability of our solutions, and we discuss possible effects on accuracy in appendix~\ref{app:resolution_test}.

These three choices for $\alpha_{\text{th}}$ in the Kippenhahn prescription also correspond to three different relationships between the timescale over which thermohaline mixing acts, $t_{\rm th}$, and the evolutionary timescale of the star, $t_{\rm{evol}}$. 
We take the thermohaline timescale to be the diffusive timescale associated with thermohaline mixing, $t_{\rm th} = d^2/D_{\rm th}$, where $d$ is the radial depth of the thermohaline zone.
In the case where $\alpha_{\text{th}} = 0.1$, 
$t_{\rm evol} \ll t_{\rm th}$, meaning the timescale for homogenization of the mixing zone is large. In the case where $\alpha_{\text{th}} = 700$, 
$t_{\rm evol} \gg t_{\rm th}$ and the homogenization timescale is short. In the intermediate case ($\alpha_{\text{th}}= 2$), the timescales are comparable; this is likewise true for the assumptions made in the Brown model, though their prescription does not involve $\alpha_{\text{th}}$. 
Given the range of mixing timescales probed, these models should confer some insight as to how (or whether) the inferred fluid parameters, including the reduced density ratio $r$, depend on both the input mixing timescale and the stellar parameters.


\subsection{Method for Extracting $r$}
We wish to extract the reduced density ratio $r$. As described in Sections \ref{intro:subsec:fluids} and \ref{sec:formalism}, $r$ characterizes the stability of the mean molecular weight gradient in the thermohaline region above the burning shell. It measures the region's tendency to mix, not its mixing efficiency.
We define a selection criterion that averages over many mass shells and many evolutionary time steps to ensure that our measured values of $r$ are representative of thermohaline mixing during the relevant evolutionary regime.

To measure $r$ in our MESA simulations, we first restrict to the appropriate evolutionary phase. We exclude all models for which MESA does not detect thermohaline mixing within $m_{\rm max} \leq m_i < 1.1 m_{\rm max}$, where $m_i$ is the mass coordinate of the $i$th mass shell and $m_{\rm max}$ is the mass coordinate coinciding with the instantaneous peak of the nuclear energy generation.
The thermohaline zone extends from a maximum mass coordinate $m_{\rm heavy}$ to a minimum mass coordinate $m_{\rm light}$ with stratification $\Delta m = m_{\rm heavy} - m_{\rm light}$.
We exclude the first 21 models in which the thermohaline zone spans at least 10 mass shells.
We then compute the evolution of $\Delta m$ of the $j$th model, $\delta m^j = \Delta m^{j} - \Delta m^{j-1}$ for model $j$ and the previous 20 models and compute $\langle \delta m \rangle = (1/20)\sum_{j=-20}^0 \delta m^j$ to determine whether the mass of the thermohaline region has evolved appreciably over the past several timesteps. We expect $\langle \delta m \rangle$ to be relatively large when the thermohaline zone is developing and small when it is in a relatively steady state.
We then measure $\epsilon = |\langle \delta m \rangle / \mathrm{max}(\Delta m^j)|$. If $\epsilon < 5 \times 10^{-3}$, we consider the model to have reached a steady state of thermohaline mixing (classified as ``good'' or ``stable''), at which point we compute $r$.

To compute the reduced density ratio $r = (R_0 - 1)/(\tau^{-1} - 1)$, we take the volume average $\bar{r} = \sum r_i dV_i / \sum dV_i$ over a subset of mass bins $i$ of the thermohaline zone. We volume-average the reduced density ratio $r$ over the mass range bounded by $m_{\rm{heavy}} + 0.1\Delta m  < m_i \leq m_{\rm heavy} + (0.1 + 1/3)\Delta m$.
In the volume average, we set the volume element $dV_i = 4\pi r_i^2 \Delta r_i$ and perform integration using the composite trapezoidal rule as implemented in \texttt{NumPy} \citep{numpy}.
We stop extracting $r$ after we have collected measurements over 1000 models, which captures the behavior of the saturated thermohaline zone and its eventual merging with the convective envelope. For each stellar evolution simulation, we report the median of the volume-averaged $r$ over all of the stable models in which measurements were taken. Results are discussed in terms of the logarithm of this quantity, $\log_{10} r$.

A movie demonstrating the evolution of a thermohaline front and the reduced density ratio selection algorithm is available in Appendix~\ref{app:movie}.

\section{Results from Numerical Experiments }
\label{sec:mesa_results}
The reduced density ratio $r$ is a measure which relates the star's stable thermal stratification to the destabilizing inverted mean molecular weight stratification in the thermohaline zone.
Regions with small $r$ ($r < 1$) are destabilized by a mean molecular weight inversion and will exhibit thermohaline mixing.
Regions with large $r$ ($r \geq 1$) have a mean molecular weight inversion that is too weak to overcome the stable thermal stratification, and thus cannot drive thermohaline instability.
Note that $r$ alone does not determine the \emph{efficiency} of thermohaline mixing.
The efficiency is determined by another function, $D_{\rm{th}}(r)$, such as the prescriptions laid out in  Sec.~\ref{sec:parameterizations}.
Most prescriptions show $D_{\rm{th}}(r)$ decreasing with increasing $r$, but some new prescriptions \citep[e.g.,][]{harrington} suggest that $\Dth(r)$ may increase with increasing $r$.

Our goal is to measure how $r$ varies with stellar mass and metallicity in RGB thermohaline zones.
Put differently, we want to understand how much nuclear reactions in the hydrogen burning shell destabilize the mean molecular weight gradient, and how large that instability is compared to the thermally stable background.
Getting a robust measure of the average value of $r$ in a thermohaline zone is difficult, because the thermohaline instability mixes the fluid, changing the mean molecular weight profile, and ultimately changing $r$ in a manner according to the prescription $\Dth(r)$ used.
In order to understand both the ``natural'' value of $r$ that the stellar model inherits (i.e.~in the absence of significant mixing) and how choice of mixing models affect $r$, we run four grids of models.
We run two grids where the thermohaline mixing timescale and the evolutionary timescale are comparable (Brown, Kippenhahn with $\alpha_{\rm{th}} = 2$).
We run a third grid where thermohaline mixing is fast compared to evolution (Kippenhahn, $\alpha_{\rm th} = 700$), from which we can understand how rapid mixing affects the value of $r$.
Finally, we run a fourth grid where evolution is fast compared to mixing (Kippenhahn, $\alpha_{\rm{th}} = 0.1$), which allows us to probe $r$ when mixing does not appreciably modify the composition profile.

\begin{figure*}
    \centering
    \includegraphics[width=\textwidth]{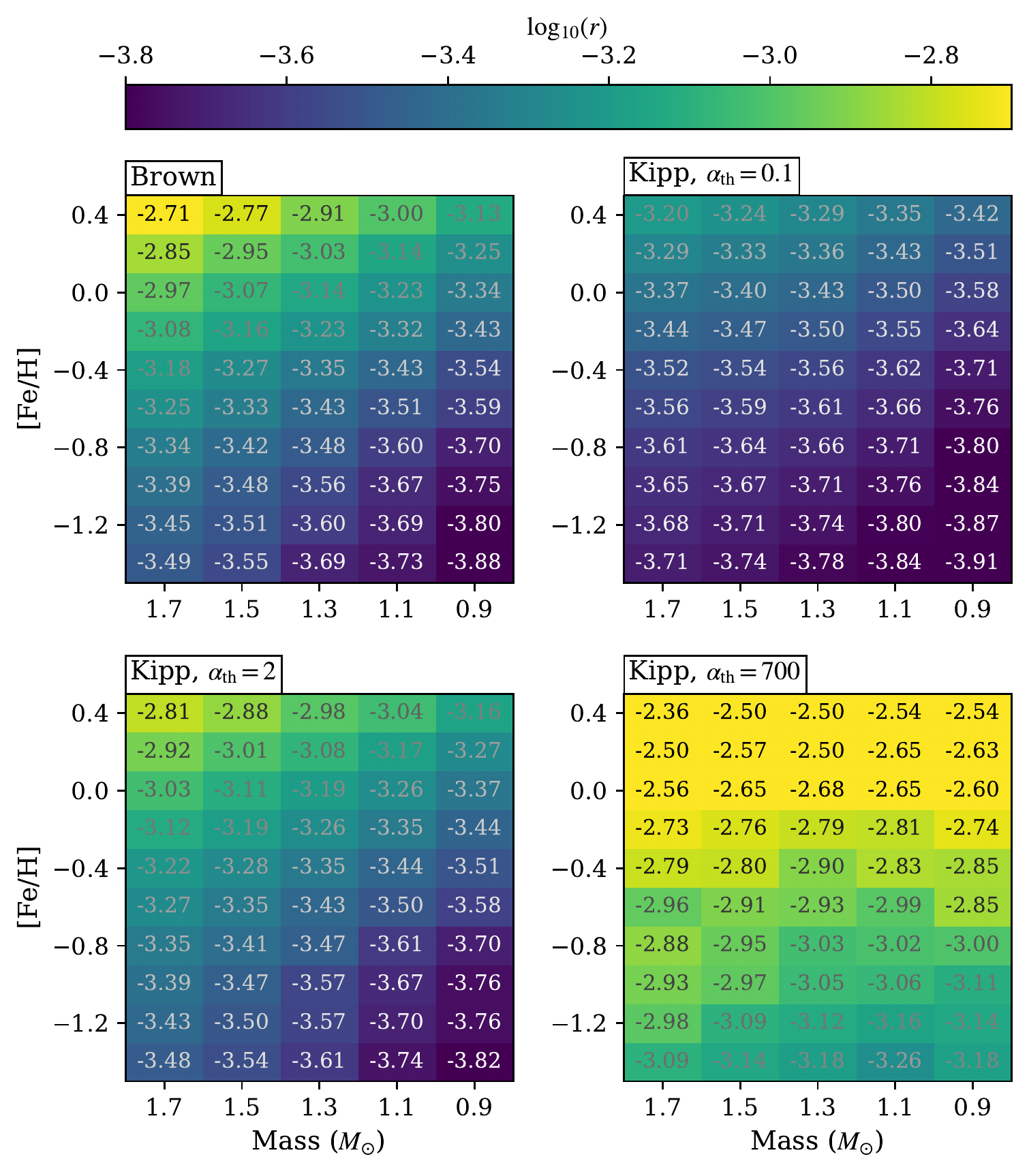}
    \caption{The reduced density ratio $\log_{10} r$ is extracted as discussed in Section \ref{sec:mesa_experiment} for four grids of stellar models with differing prescriptions for thermohaline mixing. 
    Results for $\log_{10} r$ are shown as a function of stellar mass and metallicity [Fe/H], with high values of $\log_{10} r$ in brighter colors (yellow) and low values of $\log_{10} r$ in darker colors (purple). 
    The model name and mixing efficiency, $\alpha_{\text{th}}$ (where applicable) constitute the physical configuration and are indicated in the panel labels.}
    \label{fig:mesa_r_spread}
\end{figure*}

Results from these four physical configurations are compared in Fig.~\ref{fig:mesa_r_spread}: the upper left panel shows results from the Brown model; the remaining three show results from the Kippenhahn prescription with $\alpha_{\text{th}}$ varying as indicated. The reduced density ratio $\log_{10} r$ is shown as a function of mass and metallicity and indicated on the color bar and grid labels.

In all cases, the most notable trend is that $\log_{10} r$ decreases along the diagonal from high masses and metallicities (upper left) to low masses and metallicities (lower right). There is particularly high qualitative similarity between the Brown model and Kippenhahn model with $\alpha_{\text{th}} = 2$, which correspond to similar thermohaline mixing timescales. The case with the lowest mixing parameterization is the Kippenhahn $\alpha_{\text{th}} = 0.1$ case, and there the span of $\log_{10} r$ values is smallest. We also note that, unlike in the other three cases, $\log_{10} r$ does not scale precisely monotonically with either mass or [Fe/H] in the Kippenhahn $\alpha_{\text{th}} = 700$ case. This makes sense, because this is the case where mixing is most efficient, and so $r$ is measuring the results of the mixing prescription rather than e.g., the rate at which destabilizing $^{3}\rm{He}$ is burned. 

While there is no clear relationship between the spread of $\log_{10} r$ values observed when using the Kippenhahn prescriptions and the values of $\alpha_{\text{th}}$ adopted in each, there is a clear relationship between the median values of $\log_{10} r$ and $\alpha_{\text{th}}$: the reduced density ratios are larger when mixing is highly efficient (i.e. $t_{\mathrm th}\ll t_{\text{evol}}$). 
This is consistent with $\Dth(r)$ in Eqn.~\ref{eq:Dth-kipp-MESA}; mixing increases $r$, which in turn decreases the mixing efficiency $\Dth$ in the Kippenhahn model, and so an equilibrium between the destabilizing source and the mixing is reached at a higher value of $r$ (and a corresponding lower efficiency in that prescription).

Most importantly, the overall behavior of $\log_{10} r$ as a function of mass and [Fe/H] is consistent regardless of the thermohaline parameterization adopted.
The robust \emph{trend} across 1D thermohaline mixing model assumptions suggests that $r$ may be useful to compare to physical data sets.

Note that $r$ is a measurable parameter with a robust trend in metallicity and mass, $r(M, [\mathrm{Fe/H}])$. 
In the next section, we will measure changes in [C/N] as a function of the same parameter space $\Delta [\mathrm{C/N}](M, [\mathrm{Fe/H}])$, which is a proxy for $\Dth(r)$ integrated over time.
Then we will compare $\Delta [\mathrm{C/N}]$ and $r$.
A monotonic trend between these quantities suggests a simple monotonic relationship $\Dth(r)$, and could be used to reduce the parameter space from a two-dimensional space (mass, metallicity) to a one-dimensional space (reduced density ratio, $r$).
Furthermore, the direction of the trend can be used to validate thermohaline mixing prescriptions.
For example, the purple HG19 line and the green Kippenhahn line in Fig.~\ref{fig:parameterization_compare} would produce different trends in this comparison of observational mixing vs.~$r$ because of the opposite manner in which mixing efficiency depends on $r$.

\section{Observed Mixing Signatures }
\label{sec:obs}
As discussed in Section \ref{sec:parameterizations}, we are trying to distinguish between different models of thermohaline mixing based not on the amount of mixing they predict, but on their trends as a function of the reduced density ratio $r$. As discussed in Section \ref{sec:mesa_results}, this requires stars of a wide range of masses and metallicities. It is therefore quite convenient that 
modern spectroscopic surveys have recently begun collecting measurements of mixing diagnostics for large samples of stars whose masses are also well constrained. 
%
We choose for this work to use the carbon-to-nitrogen ratios, [C/N], measured from the Apache Point Galactic Evolution Experiment \citep[APOGEE, ][]{Majewski2017}. APOGEE is a Sloan Digital Sky Survey III and IV \citep{Blanton2017} project using the 2.5 meter Sloan Telescope \citep{Gunn2006} and the APOGEE spectrograph \citep{Wilson2019} to obtain medium resolution (R $\sim$ 22,500) spectra of large numbers of stars across the galaxy \citep{Zasowski2017, Beaton2021,Santana2021}. These spectra are homogeneously reduced and analyzed using the ASPCAP pipeline \citep{Nidever2015, Zamora2015, GarciaPerez2016} and the resulting stellar parameters are then calibrated using asteroseismic, cluster, and field data \citep{Holtzman2015,Holtzman2018, Jonsson2020}. We choose to use the APOGEE data because this calibration work has already been done and an asteroseismic overlap sample is already available to provide stars with precise and accurate masses, though 
similar work could likely be done with, for example, the lithium abundances measured by the GALAH survey \citep{buder2019} or the \ctwelvecthirteen data estimated from the APOGEE data using the Brussels Automatic Code for Characterizing High accUracy Spectra \citep[BACCHUS,][]{Masseron2016_BACCHUS} pipeline (C. Hayes, submitted). 

\begin{figure}[!tb]
\begin{center}
\includegraphics[width=9cm,clip=true, trim=0.5in 0in 0in 0in]{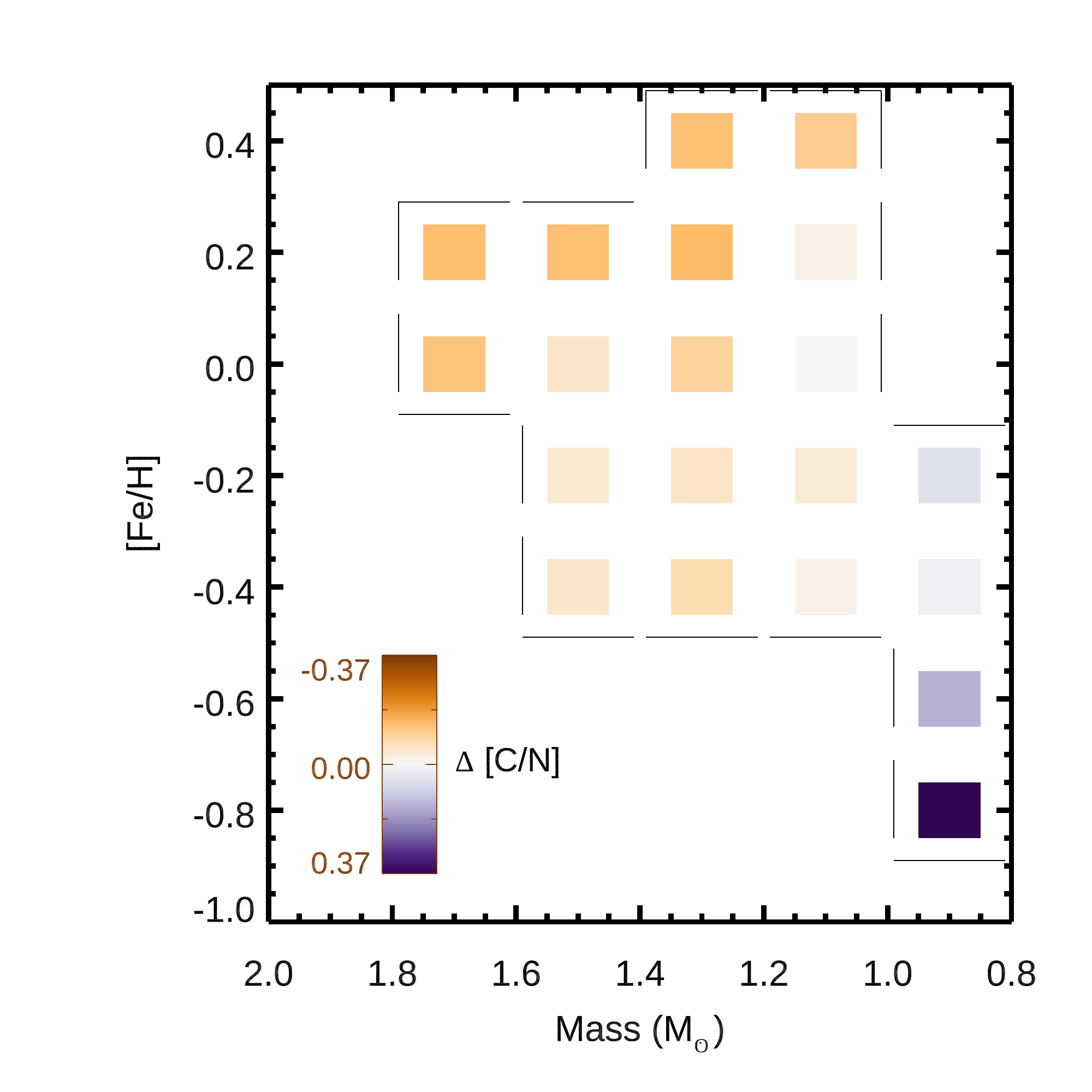}
\caption{
The difference in average [C/N] (indicated by box color, with negative values in orange and positive values in purple) between stars significantly below the RGB bump and those significantly above the bump is shown as a function of stellar mass and metallicity. The gradient is consistent with previous work, with lower mass, lower metallicity stars having more extra mixing (purple), but there is clearly an unphysical `unmixing' trend (orange) that needs to be removed (see text). We highlight only bins with a sufficient number of stars both below and above the bump. }
\label{fig:obssquare}
\end{center}
\end{figure}

We also note that the evolution of [C/N] is in some ways simpler for these low-mass stars than other mixing diagnostics. Unlike lithium, its abundance at the surface does not change significantly during the main sequence due to the effects of rotational and other mixing processes \citep{Iben1967}. Its initial ratio seems to be somewhat metallicity dependent \citep{Shetrone2019}, with higher values at lower metallicity. As stars reach the first dredge up, there is a strong, rapid, mass-- and metallicity--dependent change in the surface [C/N] ratio \citep{Salaris2015, MasseronGilmore2015, Martig2016, Ness2016, Spoo2022}. The [C/N] ratio at the surface then remains constant until stars reach the red giant branch bump, after which there seems to be another rapid drop in the [C/N] ratio, particularly in stars of low metallicity \citep[e.g.][]{Gratton2000,Shetrone2019}; it is this drop that has been associated with thermohaline mixing.  For stars of particularly low metallicities, there are some suggestions of Upper RGB extra mixing \citep{Shetrone2019}, but this is not well motivated theoretically and is distinct from the processes we are discussing here.

To estimate the amount of extra mixing in these stars near the bump---which thermohaline models suggest should correlate with the mixing coefficient $\Dth$ described above---\citet{Shetrone2019} estimated the drop in [C/N] just above the red giant branch bump. Their work used $\alpha$-element enhanced, and therefore old and low-mass ($\sim$0.9 \msun), first ascent red giant branch stars and binned them in bins of 0.2 dex in metallicity. The location of the red giant branch bump was identified empirically as an overdensity of stars at a particular surface gravity in each bin. They then identified the $\log g$ regime around the red giant branch bump and fit a hyperbolic tangent function to measure the location and size of the drops in the [C/N] ratio. For simplicity, we have reproduced their results in Table \ref{tab:obsdata}. 

\begin{table*}[tb]
\begin{center}
\caption{Observed extra mixing drops in bins of mass and metallicity, corrected for the  0.1456 dex of unmixing observed that we assume is due to systematic errors. We also include  the reduced density ratios  calculated for each of these bins using the variety of models discussed in Section \ref{sec:mesa_results}.}
\begin{tabular}{rrrrrrrr}
\hline
\multicolumn{1}{l}{M} & \multicolumn{1}{l}{[Fe/H]} & \multicolumn{1}{l} {$\Delta$[C/N]$_{\rm APK, cor}$} & {$\Delta$[C/N]$_{\rm Shet, cor}$}  & \multicolumn{1}{l}{$r_{\rm Brown, 1}$} & \multicolumn{1}{l}{$r_{\rm Kip, 0.1}$} & \multicolumn{1}{l}{$r_{\rm Kip, 2}$} & \multicolumn{1}{l}{$r_{\rm Kip, 700}$} \\ \hline \hline
0.9 & -1.4 & ... & \multicolumn{1}{r}{0.73} & 0.00013 & 0.00012 & 0.00015 & 0.00066 \\ 
0.9 & -1.2 & ... & \multicolumn{1}{r}{0.67} & 0.00016 & 0.00013 & 0.00017 & 0.00073 \\ 
0.9 & -1.0 & ... & \multicolumn{1}{r}{0.48} & 0.00018 & 0.00014 & 0.00017 & 0.00078 \\ 
0.9 & -0.8 & 0.52 & \multicolumn{1}{r}{0.36} & 0.00020 & 0.00016 & 0.00020 & 0.00099 \\ 
0.9 & -0.6 & 0.29 & \multicolumn{1}{r}{0.27} & 0.00026 & 0.00018 & 0.00026 & 0.00140 \\ 
0.9 & -0.4 & 0.16 & \multicolumn{1}{r}{0.21} & 0.00029 & 0.00019 & 0.00031 & 0.00143 \\ 
1.1 & -0.4 & 0.13 & ... & 0.00037 & 0.00024 & 0.00036 & 0.00147 \\ 
1.3 & -0.4 & 0.07 & ... & 0.00045 & 0.00027 & 0.00045 & 0.00126 \\ 
1.5 & -0.4 & 0.10 & ... & 0.00054 & 0.00029 & 0.00053 & 0.00160 \\ 
0.9 & -0.2 & 0.20 & ... & 0.00037 & 0.00023 & 0.00036 & 0.00181 \\ 
1.1 & -0.2 & 0.11 & ... & 0.00048 & 0.00028 & 0.00045 & 0.00156 \\ 
1.3 & -0.2 & 0.09 & ... & 0.00059 & 0.00032 & 0.00055 & 0.00161 \\ 
1.5 & -0.2 & 0.10 & ... & 0.00069 & 0.00034 & 0.00065 & 0.00173 \\ 
1.1 & 0.0 & 0.14 & ... & 0.00059 & 0.00032 & 0.00054 & 0.00224 \\ 
1.3 & 0.0 & 0.05 & ... & 0.00072 & 0.00037 & 0.00064 & 0.00210 \\ 
1.5 & 0.0 & 0.09 & ... & 0.00085 & 0.00039 & 0.00078 & 0.00225 \\ 
1.7 & 0.0 & 0.02 & ... & 0.00107 & 0.00042 & 0.00093 & 0.00275 \\ 
1.1 & 0.2 & 0.12 & ... & 0.00072 & 0.00037 & 0.00068 & 0.00226 \\ 
1.3 & 0.2 & 0.00 & ... & 0.00094 & 0.00043 & 0.00083 & 0.00317 \\ 
1.5 & 0.2 & 0.01 & ... & 0.00112 & 0.00047 & 0.00098 & 0.00267 \\ 
1.7 & 0.2 & 0.00 & ... & 0.00142 & 0.00051 & 0.00119 & 0.00318 \\ 
1.1 & 0.4 & 0.04 & ... & 0.00100 & 0.00045 & 0.00092 & 0.00289 \\ 
1.3 & 0.4 & 0.01 & ... & 0.00124 & 0.00052 & 0.00104 & 0.00319 \\ 
 \hline
\end{tabular}
\label{tab:obsdata}
\end{center}
\end{table*}

We add to their analysis a sample of higher metallicity stars with asteroseismic masses from the APOGEE-Kepler overlap sample \citep[APOKASC,][]{Pinsonneault2014, Pinsonneault2018}. We do this because, according to our analysis in Section \ref{sec:mesa_results}, higher mass, higher metallicity stars probe larger values of the reduced density ratio, $r$. 
We first bin the stars in mass (0.2 \msun) and  metallicity (0.2 dex). For consistency with \citet{Pinsonneault2018} and \citet{Shetrone2019}, we use the Data Release 14 \citet{DR14} carbon and nitrogen abundances. We note however that while the abundance scale seems to shift between releases, the rank ordering does not change very much \citep{Spoo2022}, which means that the conclusions of this analysis are not strongly affected by the choice of Data Release or seismic parameters.

Unlike in the \citet{Shetrone2019} analysis (e.g. their Figure 2), there is not a sufficient number of stars near the bump in each bin to detect and measure the extra mixing directly in the asteroseismic sample. Instead, we define a `pre-mixing' bin of stars between \logg\ of 3.4 and 2.8 dex whose oscillations have identified them as first ascent red giants \citep{Elsworth2019}, as well as a `post-mixing' bin of RGB stars with surface gravities between 2.3 and 1.0 dex. We then compute the average [C/N] of stars in each of the pre-mixing and post-mixing bins. If both bins had at least three stars, then the difference between the pre-mixing and post-mixing average [C/N] is plotted in Figure \ref{fig:obssquare}. Because of the calibration and choices in the analysis pipeline
\citep[see e.g.][]{Holtzman2018,Jonsson2020, vsmith_apogee_dr16_2021}, the scale of the abundances, particularly for carbon and nitrogen, is somewhat uncertain.
There sometimes exist small trends with surface gravity and temperature that are not fully removed in the calibration process. This is notable in our measurement results here; in the highest mass, highest metallicity bins, we formally measure `unmixing' near the red giant branch bump, i.e. an increase rather than a decrease in the average [C/N] near the red giant branch bump, which is inconsistent both with theoretical expectations and with measurements from other sources. Following discussions with the APOGEE team (C. Hayes, private communication), we have decided to correct for these effects by correcting the bin with the most `unmixing' to have 0 mixing, and subtracting that change from all of the other bins under the assumption that the systematic measurement errors are consistent for stars of similar temperatures and gravities. Such calibrations are common in the literature \citep{Holtzman2015, Buder2021}, and because we are most interested here in the trend in mixing amounts as a function of the stellar parameters, we do not expect this shift to alter the results of this analysis, but we emphasize that care should be taken by future users of this data.

\section{Results} 
\label{sec:punchline}

\begin{figure*}[!tb]
\begin{center}
\includegraphics[width=\textwidth]{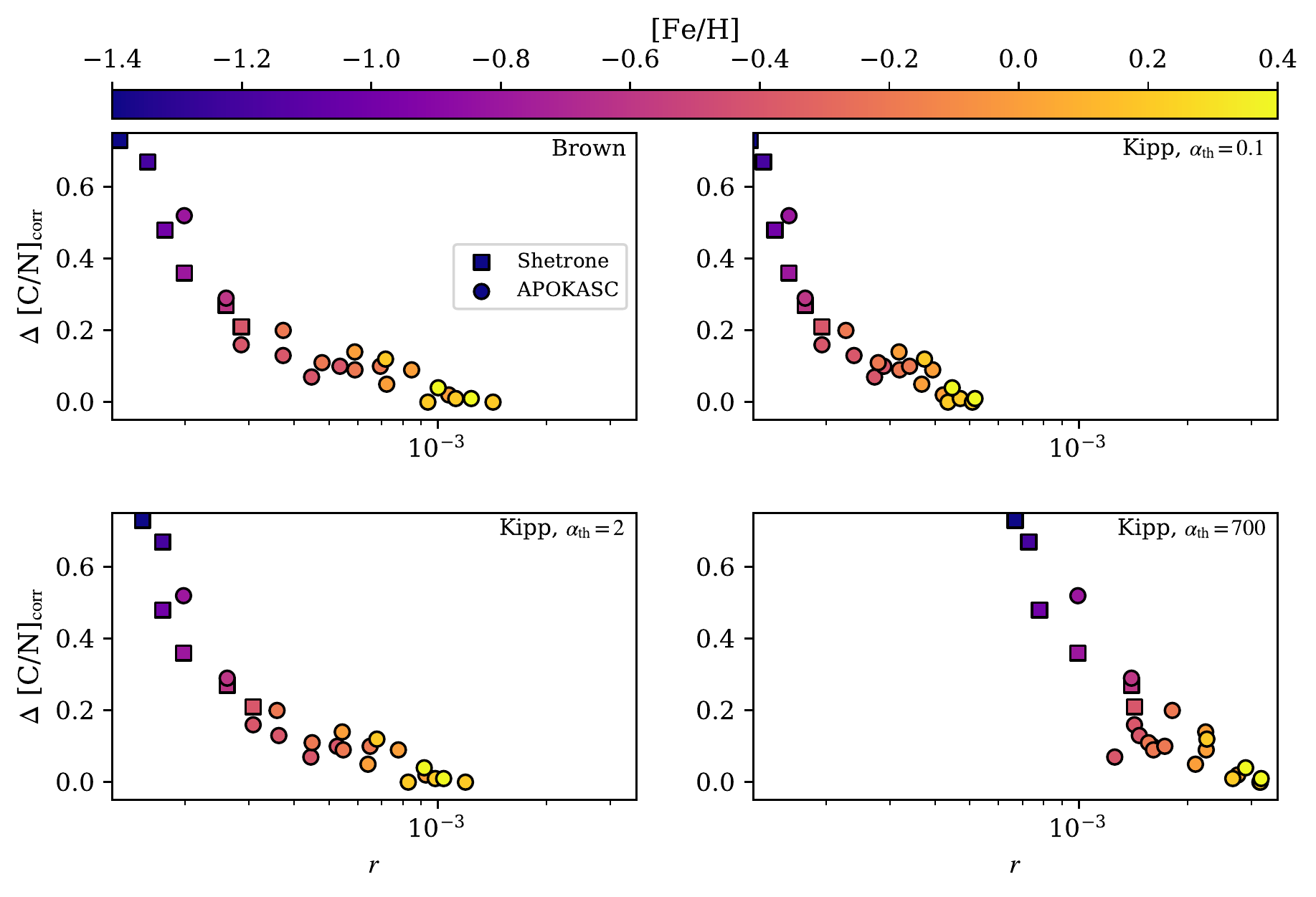}
\caption{Corrected measurements of the change in [C/N] near the red giant branch bump are compared to the reduced density ratio inferred from one-dimensional models using various thermohaline mixing prescriptions (Brown, Kippenhahn $\alpha_{\rm th}=0.1$, Kippenhahn $\alpha_{\rm th}=0.2$,Kippenhahn $\alpha_{\rm th}=0.700$). Observations are color coded by the metallicity bin of each data point. In general, there is a clear correlation between these parameters, suggesting that the observed mixing may indeed be related to
the unstable mean molecular weight gradient. Mixing and the reduced density ratio $r$ are inversely correlated, which is consistent with hydrodynamic thermohaline prescriptions. 
\label{Fig:punchline}
}
\end{center}
\end{figure*}

Given the measured amounts of mixing described in Section \ref{sec:obs} and the reduced density ratios $r$ computed for stars of various masses and metallicities described in Section \ref{sec:mesa_results}, it is now possible to compare the observations to the predictions of various thermohaline models from Section \ref{sec:formalism} and to assess whether the observed mixing is qualitatively consistent with any such theoretical prescription.
We show in Figure \ref{Fig:punchline} the corrected changes in [C/N] compared to the inferred reduced density ratios on axes analogous to those of Figure \ref{fig:parameterization_compare}, where the y axis represented the rate of mixing. 
The four panels correspond to the four modeling configurations described in Section \ref{sec:mesa_experiment}. 

\begin{figure*}[!tb]
\begin{center}
\includegraphics[width=\textwidth]{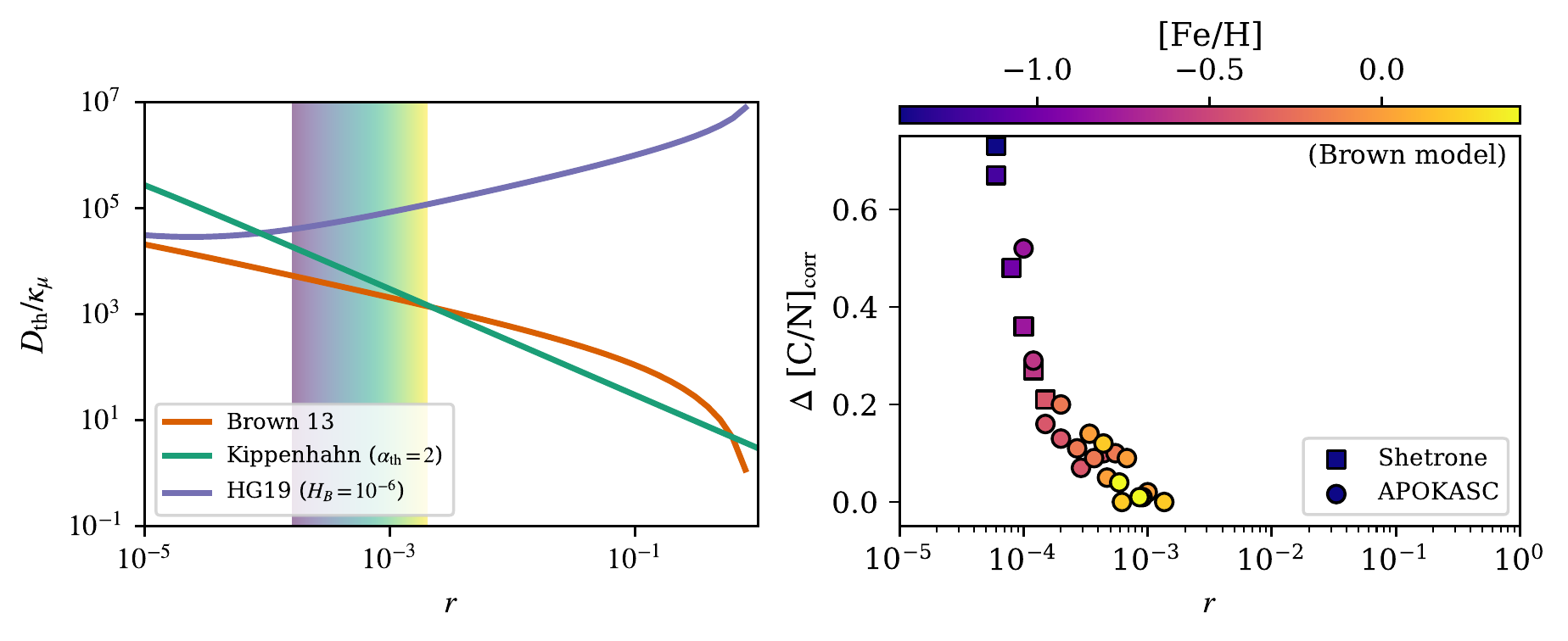}
\caption{Left: A reproduction of Figure \ref{fig:parameterization_compare} showing the predicted rate of mixing versus the reduced density ratio in various prescriptions of thermohaline mixing, including hydrodynamic (orange, green) and magnetohydrodynamic (purple) models. Right: The observed extra mixing near the red giant branch bump as a function of the reduced density ratio inferred from one dimensional stellar evolution models. While the conversion from the change in a mixing diagnostic to the fluid mixing rate is not trivial, and therefore we do not attempt it here, we note that the observed mixing amounts are strongly negatively correlated with $r$, with stars probing on average a relatively narrow range of the regime formally unstable to thermohaline mixing. }
\label{Fig:compare}
\end{center}
\end{figure*}

We first note that the observed trends are not strongly sensitive to the assumed 1D mixing model: mixing decreases with increasing reduced density ratio $r$ regardless of parameterization.
Besides this, there are similarities and differences between the data and the theoretical predictions, which are reproduced in Figure \ref{Fig:compare} for comparison. A key finding is that \textit{the observed mixing is strongly correlated with the fluid parameter $r$} as predicted; this is true for stars with different masses and metallicities but similar reduced density ratios.
This parameter therefore reduces a 2D (mass, metallicity) parameter space to one.
We observe a decrease in the amount of mixing as the density ratio increases, which is consistent with standard 1D prescriptions of thermohaline mixing but inconsistent with the prescription from \citet{harrington}, which was informed by magnetohydrodynamic simulations. 
We also find that the range of average reduced density ratios probed by the observational data we have available here is much smaller than the range of density ratios simulated by and studied within the theoretical 3D fluid dynamics community \citep[e.g.][]{brown_etal_2013}, although we note that the full range of ratios do appear in each individual simulation (see Appendix \ref{app:movie}).

\section{Conclusions }
\label{sec:conclusions}
Thermohaline mixing has long been considered the most likely candidate for explaining the evolution of the surface chemistry of low--mass upper red giant branch stars. In this analysis, we have shown that:
\begin{enumerate}
    \item prescriptions informed by three-dimensional simulations of the thermohaline instability suggest that mixing rates should strongly depend on the reduced density ratio, $r$, but the shape of the correlation varies from model to model; 
    
    \item one-dimensional stellar evolution models suggest that the average reduced density ratio, $r$, should vary as a function of mass and metallicity, with stars of lower masses and lower metallicities having smaller $r$;
    
    \item one-dimensional stellar evolution models suggest that the average reduced density ratio, $r$, is not strongly dependent on the assumed parameterized thermohaline mixing model (e.g. Brown vs Kippenhahn), and regardless of the choice of $\alpha_{\rm th}$ in the case of Kippenhahn; 
    
    \item one-dimensional stellar evolution models suggest that the average reduced density ratio, $r$, occupies only a small range ($10^{-4}<r<10^{-3}$) of the parameter space formally unstable to thermohaline mixing, $0<r<1$;
    
    \item observations suggest that the amount of mixing is strongly correlated with the reduced density ratio, $r$, in a way that qualitatively agrees with predictions from three dimensional simulations;
    
    \item observations indicate that the mixing rate and reduced density ratio, $r$, are inversely correlated, a finding that is consistent with 
    currently available 1D prescriptions informed by 3D simulations, but not consistent with the prescription put forth by \citet{harrington} with fixed magnetic parameter $H_B$.
    
    \end{enumerate}

Most importantly, we find that our proposed framework for connecting observations of extra mixing in red giants to the fluid simulations of the thermohaline instability through the medium of one dimensional stellar evolution models is feasible and robust. Whereas previous work has focused on calibrating individual thermohaline mixing prescriptions against observations, this framework allows for comparison between different mixing models themselves. 
It motivates more rigorous exploration into whether red giant branch extra mixing should be associated with the thermohaline instability, and it will facilitate the translation of observational information into theoretical simulations. As observational data sets improve, abundance measurements will only become more capable of constraining stellar interior mixing, the relevant fluid parameters, and potentially even the magnetic fields in these regions.

However, the present study is largely a proof of concept;
there is room for significant development in all aspects of this method. From an observational perspective, we have so far only considered the change in [C/N] as an observational mixing diagnostic, and only in a fairly restricted set of stars. Looking at higher or lower masses, using a variety of mixing diagnostics, or probing stars including globular clusters, binary stars, and dwarf galaxies---where the base composition and mixing history may be different---could all be illuminating expansions of this work. There is also the potential to look at the timing of the extra mixing, as well as mixing rate as a function of time. Further, it may be possible to investigate whether mixing can be connected to the sorts of asteroseismic diagnostics that probe the interior structure of a star, including its density profile \citep{KjeldsenBedding1995}, chemical discontinuities \citep{Verma2017}, internal rotation \citep{Gehan2018}, and magnetic fields \citep{Bugnet2021}. 

On the modeling side, we have explored a coarse but reasonable range of possible thermohaline conventions that could have impacted our conclusions, but this parameter exploration was certainly not exhaustive. It is worthwhile and necessary to test whether different 1D modeling assumptions impact the direction of this trend, particularly in the case of other mixing--related physical assumptions. Key variations in this regard include the choice of prescription for the Mixing Length Theory (MLT) formalism and value for the mixing length parameter, $\alpha_{\text{MLT}}$, the treatment of convective boundary mixing and convective overshoot, and the choice of atmospheric surface boundary conditions---all of which are well known to affect thermodynamic quantities in the regime we study here \citep{tayar2017, Joyce2018a, Joyce2018b, viani2018}. For similar thermodynamic reasons, it is also important to explore more extreme metallicity regimes, as the global metal content dictates the behavior of \gradmu.
Further, the need to introduce rotational mixing alongside thermohaline mixing in 1D stellar models to achieve the desired observational reproductions is likewise well established in the literature (c.f.~\citealt{CharbonnelLagarde2010}), making the consideration of rotational effects an obvious candidate for future theoretical work.

Finally, while previous and ongoing work has endeavored to find a theoretical justification for why thermohaline mixing should be so efficient that it accounts for the entire amount of extra mixing observed after the RGB bump (by adding e.g.~magnetic fields or rotation), this work sets a new target for theoretical efforts: in order for thermohaline mixing to be the primary source of extra mixing, not only should there be a mixing prescription that agrees with 3D simulations and also reproduces the \textit{amount} of extra mixing, but such a prescription should also reproduce \textit{trends} in this extra mixing as a function of fundamental stellar parameters like mass and metallicity.
For instance, while \citet{harrington} demonstrate that magnetic fields of moderate strength can dramatically enhance the efficiency of thermohaline mixing---plausibly to levels that explain the full amount of extra mixing observed after the RGB bump---this work shows that their mixing prescription does not predict trends in mixing rate as a function of the density ratio that are consistent with observations.

In short, this paper has demonstrated the viability of comparing observed signatures of extra mixing on the red giant branch to the predictions of models informed directly by fluid simulations, and it has done so in the framework of parameters used in those simulations. We therefore anticipate and welcome future explorations that will produce better understanding of 
whether extra mixing should indeed be associated with the thermohaline instability and 
how observations of real stars---which probe a fluid regime well outside the regime we are currently able simulate--- can provide constraints on the physics of that instability.

\begin{acknowledgements}
The first four authors of this manuscript contributed equally to this work. 
\begin{itemize}
\item[] A.~Fraser contributed the majority of text and was responsible for Secs.~2 and 3.
\item[] M. Joyce wrote and tested the MESA modeling templates, including \texttt{inlists}, evolutionary and structural output, and \texttt{run\_star\_extras} functionality, wrote the data analysis and parameter extraction scripts in collaboration with E.H. Anders (Python), and contributed text.
\item[] E.H. Anders computed and analyzed stellar structure models, created Figs.~2 and 4-7, and contributed text.
\item[] J. Tayar was responsible for all analysis related to observations, constructed the initial manuscript template, and contributed text. 
\end{itemize}
Author order within this set was decided according to height in heels, descending.

The authors thank C. Hayes, A. Jermyn, and M. Pinsonneault for helpful discussions that contributed to this work and T. White for providing the script that translates between [Fe/H] and Z. The authors acknowledge the thermohaline working group at the KITP program ``Probes of transport in stars'' for fruitful discussion that led to the conception of this idea. The authors likewise thank the Space Telescope Science Institute for providing accommodation and infrastructure support during the week-long meeting during which the majority of this paper was written.
AF thanks Pascale Garaud for many discussions of thermohaline mixing which helped improve his understanding of the theory of that field. 
MJ thanks John Bourke for proofreading and helpful discussion of numerics.
EHA thanks MJ for her mentorship in teaching him how to run and interact with MESA models.
Support for this work was provided by NASA through the NASA Hubble Fellowship grant No.51424 awarded by the Space Telescope Science Institute, which is operated by the Association of Universities for Research in Astronomy, Inc., for NASA, under contract NAS5-26555. JT also acknowledges support from 80NSSC20K0056. AF acknowledges support from NSF Grant No.~AST-1908338, NASA HTMS grant 80NSSC20K1280, and from the George Ellery Hale Postdoctoral Fellowship in Solar, Stellar and Space Physics at University of Colorado. 
MJ acknowledges the Space Telescope Science Institute's Lasker Data Science Fellowship, the Kavli Institute for Theoretical Physics at UC Santa Barbara, and the MESA developers team.
EHA acknowledges the support of a CIERA Postdoctoral fellowship.
 The Center for Computational Astrophysics at the Flatiron Institute is
supported by the Simons Foundation.
 Computations were conducted with support from the NASA High End Computing (HEC) Program through the NASA Advanced Supercomputing (NAS) Division at Ames Research Center on Pleiades with allocation GID s2276 which is provided through NASA HTMS grant 80NSSC20K1280. 
This research was supported in part by the National Science Foundation under Grant No. NSF PHY-1748958.

\end{acknowledgements}

\software{Astropy \citep{Astropy,Astropy_2018}, Matplotlib \citep{Matplotlib}, NumPy \citep{numpy}, SciPy \citep{2020SciPy-NMeth}}

\facilities{Du Pont (APOGEE), Sloan (APOGEE)} 

\bibliographystyle{aasjournal}
\bibliography{ms.bib, library.bib, library2.bib, thermohaline.bib, mesa.bib, Joyce_bibliography_4.12.22}

\appendix

\section{MESA}
\label{app:mesa}
The MESA EOS is a blend of the OPAL \citep{Rogers2002}, SCVH
\citep{Saumon1995}, FreeEOS \citep{Irwin2004}, HELM \citep{Timmes2000},
PC \citep{Potekhin2010}, and Skye \citep{Jermyn2021} EOSes.

Radiative opacities are primarily from OPAL \citep{Iglesias1993,
IglesiasRogers1996}, with low-temperature data from \citet{Ferguson2005}
and the high-temperature, Compton-scattering dominated regime by
\citet{Poutanen2017}.  Electron conduction opacities are from
\citet{Cassisi2007}.

Nuclear reaction rates are from JINA REACLIB \citep{Cyburt2010}, NACRE \citep{Angulo1999} and
additional tabulated weak reaction rates \citet{Fuller1985, Oda1994,
Langanke2000}.  Screening is included via the prescription of \citet{Chugunov2007}.
Thermal neutrino loss rates are from \citet{Itoh1996}.

We create 1D stellar models and evolve them from the pre-main sequence until roughly the end of hydrogen shell burning.
We study stellar masses between 0.9 and 1.7 $M_{\odot}$ in steps of $0.2 M_{\odot}$. We study metallicities [Fe/H] ranging from -1.4 to 0.4 in steps of 0.2 dex. To convert from metallicity units to MESA input $Y$ and $Z$ units, we assume a linear helium enrichment law \citep[per e.g.,][sec 3.1]{choi2016} in which we adopt a Big-Bang $Y_p = 0.2485$ and $\Delta Y / \Delta Z = 1.3426$ according to Table 1 of \citet{tayar_etal_2022}. The algorithm we use to calculate $X$, $Y$, and $Z$ from these values is identical to the one used in \url{https://github.com/aarondotter/initial_xa_calculator}; we adopt the solar heavy element mixture of \citet{GrevesseSauval1998}.
The specific [Fe/H] to ($X$, $Y$, $Z$) conversions used here are shown in Table~\ref{table:feh_to_z}.

\begin{deluxetable}{c c c c}
\caption{
     Mappings between $[$Fe/H$]$ values and MESA input values of $(X, Y, Z)$.}
    \label{table:feh_to_z}
\tablehead{
\colhead{[Fe/H]} & \colhead{$X$} & \colhead{$Y$} & \colhead{$Z$}
}
\decimals
\startdata
      0.400 & 0.66214302 & 0.29971262 & 0.03814436 \\
      0.200 & 0.69253197 & 0.28229599 & 0.02517204 \\
      0.000 & 0.71318414 & 0.27045974 & 0.01635613 \\
     -0.200 & 0.72686070 & 0.26262137 & 0.01051793 \\
     -0.400 & 0.73576323 & 0.25751912 & 0.00671765 \\
     -0.600 & 0.74149343 & 0.25423501 & 0.00427157 \\
     -0.800 & 0.74515509 & 0.25213642 & 0.00270849 \\
     -1.000 & 0.74748410 & 0.25080161 & 0.00171429 \\
     -1.200 & 0.74896112 & 0.24995509 & 0.00108379 \\
     -1.400 & 0.74989606 & 0.24941926 & 0.00068468
\enddata

\end{deluxetable}

\section{Resolution testing and timestepping discussion}
\label{app:resolution_test}
\begin{figure*}[!tb]
\begin{center}
\includegraphics[width=\textwidth]{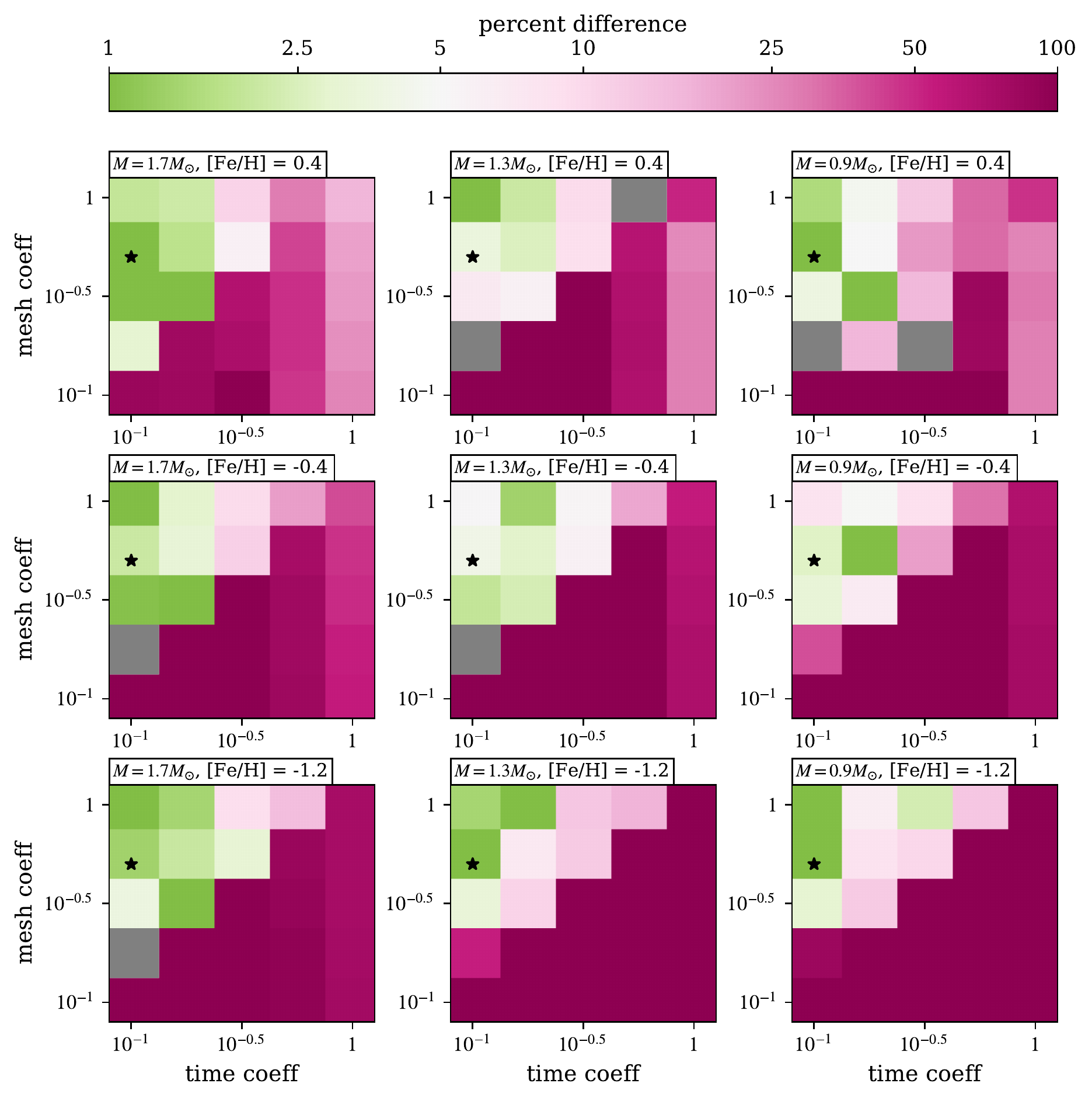}
\caption{
    We plot a 3x3 grid of colormaps corresponding to a grid of mass $M \in [0.9, 1.3, 1.7]$ and [Fe/H]$ \in [-1.2, -0.4, 0.4]$.
    At each mass and [Fe/H], we simulate a 5x5 grid of MESA models with varying spatial and temporal resolution.
    We plot in color the percent difference between the measured value of the reduced density ratio $r$ and its reference value reported in Fig.~\ref{fig:mesa_r_spread}.
    The resolution of the grids of simulations presented in Fig.~\ref{fig:mesa_r_spread} are marked by black stars.
    Cases with $r$ measurements within 5\% of the reference values are colored in green, while points with larger differences are colored pink.
    Grey pixels are simulations for which either there was a supercomputer error or the algorithm failed to identify a thermohaline zone.
    }
\label{Fig:resolution_test}
\end{center}
\end{figure*}

We performed resolution tests for models with [Fe/H] $\in \{-1.2, -0.4, 0.4\}$ and $M \in \{0.9, 1.3, 1.7\}$ using the Brown thermohaline mixing prescription.
We studied a grid of \texttt{mesh\_delta\_coeff} and \texttt{time\_delta\_coeff} values which span from 0.1 to 1.0 over five log-space steps.
This means that we evolved simulations with both spatial and temporal resolutions ranging from $1\times$ to $10\times$ the default resolutions.
For each simulation, we evolve the 1D stellar model as described in Sec.~\ref{sec:mesa_experiment}, then decrease (or increase) the spatial and temporal resolution on the red giant branch once $\log g \leq 3$.
We measure the inverse density ratio $r$ in each of these models, and in Fig.~\ref{Fig:resolution_test} we plot the absolute value of the percentage difference between that $r$ value and the reference $r_{\rm ref}$ value reported for that case in Fig.~\ref{fig:mesa_r_spread}.
We calculate the percentage difference to be $100(1 - r/r_{\rm{ref}})$.

We find that small values of the mesh coefficient (high spatial resolution) combined with large values of the time coefficient (large timesteps) result in large errors.
This occurs because the front of the thermohaline zone, and sometimes the full thermohaline zone, becomes numerically unstable, and large oscillations in $R_0$ lead to large errors in the $r$ calculation.
Furthermore, we find that when the thermohaline front is not properly numerically resolved, it does not propagate upwards in mass coordinate and so does not connect with the convective shell.
This likely has important implications for the evolution (or lack thereof) of surface abundances in these models.

    We note that \citet{lattanzio_etal_2015} report in their Eqn.~23 the Courant–Friedrichs–Lewy (CFL) criterion for diffusive processes as the timestep criterion necessary to resolve the thermohaline process in stellar models.
    This timestep criterion is required for stability when turbulent diffusivities are explicitly timestepped, but is not required for stability when diffusion is implicitly timestepped, as it is in MESA \citep[and see ][ for a more detailed discussion of CFLs and timestepping diffusive processes]{dutykh_2016}.
    Numerical stability does not imply accuracy; \citet{TayarJoyce22} studied the behavior of thermohaline mixing with $\alpha_{\rm th} = 2$ in similar stars in MESA by decreasing the timestep from well above the CFL criterion limit to well below it, and found good qualitative agreement between solutions which violated the CFL criterion and those that did not.
    This gives us some faith in our results regardless of whether they are above or below this limit.
    However, we note that this timestep limit is inversely proportional to the turbulent diffusivity $\delta t_{\rm CFL} \propto D_{\rm th}^{-1}$.
    For the Kippenhahn model (Eqn.~\ref{eq:Dth-kipp}) where $D_{\rm th} \propto \alpha_{\rm th}$, the CFL timestep limit is inversely proportional to the thermohaline mixing efficiency coefficient $\alpha_{\rm th}$ used.
    It is therefore computationally impractical for us to evolve models that do not violate the CFL criterion for the $\alpha_{\rm th} = 700$ case; it took on the order of one day for \citet{TayarJoyce22} to evolve a model that did not violate the CFL criterion with $\alpha_{\rm th} = 2$, so it would take on the order of one year for us to evolve a model with $\alpha_{\rm th} = 700$. This is prohibitively expensive and beyond the scope of this work.
    We also note that MESA is designed for convergence rather than speed, and so is roughly 10 times slower than the Monash code used by \citet{lattanzio_etal_2015}, which contributes to the unfeasibility of these calculations \citep{cinquegrana_etal_2022}.

    We include the $\alpha_{\rm th} = 700$ case in this work because it is a typical value used within the field, but we urge caution when using and interpreting the results of these simulations as they cannot be as rigorously tested as e.g., the \citet{brown_etal_2013} model.
    We have confirmed by eye that both the highest mass, highest metallicity and lowest mass, lowest metallicity simulations with $\alpha_{\rm th} = 700$ evolve in a logical manner: the thermohaline front propagates monotonically outwards and connects with the envelope CZ.

\section{Movie of thermohaline front evolution}
\label{app:movie}
\begin{figure}[!tb]
\begin{center}
\includegraphics[width=5in]{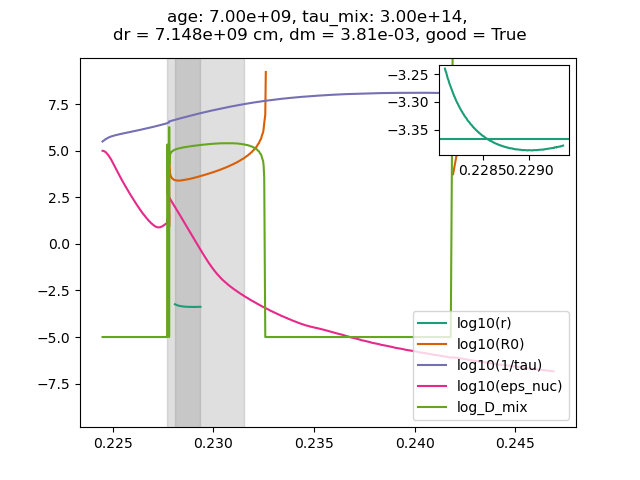}
\caption{
    We plot various stellar structure quantities (indicated in the table) vs. mass coordinate ($M / M_*$) for a $M_* = 1.1 M_\odot$ star with [Fe/H] $= -0.2$ which employs the \citet{brown_etal_2013} mixing prescription.
    We zoom in near the hydrogen burning shell, as indicated by the pink \texttt{eps\_nuc} line which shows the nuclear energy generation rate.
    The thermohaline region can be identified as the place where there is a large amount of mixing (the light green \texttt{log\_D\_mix} is large), and we shade the bulk of this region in grey.
    As described in section~\ref{sec:mesa_experiment}, we only measure $r$ over 1/3 of this region by mass, and the region over which we take this measurement is shaded in a darker grey.
    Within this region, we plot $r$ in dark green.
    A zoom-in on $log_{\rm{10}}r$ within the dark grey region is shown in the inset, and the measured value of $r$ identified by the algorithm is plotted as a dark green horizontal line.
    Additional plotted lines include $R_0$ and $1/\tau$ as described in Sec.~\ref{sec:parameterizations}.
    Various quantities are quoted in text at the top of the image, including the star's age in years, the thermohaline mixing timescale in seconds, the radial extent of the thermohaline zone in terms of both length (dr) and mass coordinate (dm), and a flag indicating that this model is identified as ``good'' by our algorithm.
    An animated version of this figure is available online in the published HTML version of this article and on YouTube at \url{https://youtu.be/XLU8aS2q5-o}.
    }
\label{Fig:movie}
\end{center}
\end{figure}

In Fig.~\ref{Fig:movie}, we plot the stellar structure vs.~mass coordinate in the simulation which employs the Brown model and has $M = 1.1M_\odot$ and [Fe/H] $= -0.2$.
We limit the x-limits of the plot to the mass coordinate energy generation peak of the hydrogen burning shell on the left $m_{\rm max}$, and to $1.1 m_{\rm max}$ on the right.
An animated version of this figure is available online in the published HTML version of this article and on YouTube at \url{https://youtu.be/XLU8aS2q5-o}.

\end{document}